\documentclass[11pt]{article}  
\def\preprint#1{%
    \thispagestyle{empty}~\newline\vspace*{-22.65mm}  
    \begin{flushright}  
    \begin{tabular}{l} #1 \end{tabular}  
    \end{flushright}  
    \vspace{1cm}}  
\def\lsim{\mathrel{\lower2.5pt\vbox{\lineskip=0pt\baselineskip=0pt  
          \hbox{$<$}\hbox{$\sim$}}}}  
\def\gsim{\mathrel{\lower2.5pt\vbox{\lineskip=0pt\baselineskip=0pt  
          \hbox{$>$}\hbox{$\sim$}}}}  
\def\real{\mathrel{\lower.0pt \hbox{$I\!\!R$}}}  
\renewcommand{\theequation}{\arabic{section}.\arabic{equation}}  

\begin{document}    
\markright{Brane Gases on K3 and Calabi-Yau Manifolds \hfil} 
 
\def\Montreal{Montr\'eal}  
\def\Quebec{Qu\'ebec}  
\title{{\small \preprint{MCGILL 01-22}} \LARGE 
BRANE GASES ON K3 \\ AND CALABI-YAU MANIFOLDS \vspace{1.0cm}}
\author{Damien A. Easson \\ [10mm]  
{\small \it   
Physics Department, McGill University,} \\
{\small \it
3600 University Street, \Montreal, \Quebec, H3A 2T8, CANADA}}  
\date{{\small \today}}  
\maketitle  
\begin{abstract}  
\baselineskip .5cm
We initiate the study of Brane Gas Cosmology (BGC) 
on manifolds with non-trivial holonomy.  Such compactifications are required
within the context of superstring theory in order to 
make connections with realistic particle physics.  We study the
dynamics of brane gases constructed from various string theories
on background spaces having a K3 submanifold.  The K3 compactifications
provide a stepping stone for generalising the model to the case of a full 
Calabi-Yau three-fold.  Duality symmetries are discussed within a cosmological
context.  Using a duality, we arrive at an N=2 theory in four-dimensions 
compactified on a Calabi-Yau manifold with SU(3) holonomy.
We argue that the Brane Gas model compactified on such spaces maintains the 
successes of the trivial toroidal compactification while greatly enhancing
its connection to particle physics.    
The initial state of the universe is taken to be a small, 
hot and dense gas of p-branes near thermal equilibrium.  The universe has no
initial singularity and the dynamics of string winding modes allow three spatial dimensions
to grow large, providing a possible solution to the dimensionality 
problem of string theory.
 
\vspace*{8mm}  
\end{abstract}  
\vfill  
\bigskip  
\centerline{\underline{E-mail:} {\tt easson@hep.physics.mcgill.ca}}  
\bigskip  
\bigskip  
\centerline{\underline{arXiv:}  
{\tt hep-th/0110225}}  
\bigskip  
\clearpage  
\def\Box{\nabla^2}  
\def\d{{\mathrm d}}  
\def\ie{{\em i.e.\/}}  
\def\eg{{\em e.g.\/}}  
\def\etc{{\em etc.\/}}  
\def\etal{{\em et al.\/}}  
\def\S{{\mathcal S}}  
\def\I{{\mathcal I}}  
\def\L{{\mathcal L}}  
\def\H{{\mathcal H}}  
\def\M{{\mathcal M}}  
\def\N{{\mathcal N}} 
\def\cP{{\mathcal P}} 
\def\R{{\mathcal R}}  
\def\K{{\mathcal K}}  
\def\W{{\mathcal W}} 

\def\mM{{\mathbf M}} 
\def\mP{{\mathbf P}} 
\def\mT{{\mathbf T}} 
\def\mR{{\mathbf R}}
\def\mS{{\mathbf S}}
\def\mX{{\mathbf X}}
\def\mZ{{\mathbf Z}}

\def\eff{{\mathrm{eff}}}  
\def\Newton{{\mathrm{Newton}}}  
\def\bulk{{\mathrm{bulk}}}  
\def\brane{{\mathrm{brane}}}  
\def\matter{{\mathrm{matter}}}  
\def\tr{{\mathrm{tr}}}  
\def\normal{{\mathrm{normal}}}  
\def\implies{\Rightarrow}  
\def\half{{1\over2}}  
\newcommand{\da}{\dot{a}}
\newcommand{\db}{\dot{b}}
\newcommand{\dn}{\dot{n}}
\newcommand{\dda}{\ddot{a}}
\newcommand{\ddb}{\ddot{b}}
\newcommand{\ddn}{\ddot{n}}
\def\be{\begin{equation}}
\def\ee{\end{equation}}
\def\bea{\begin{eqnarray}}
\def\eea{\end{eqnarray}}
\def\bs{\begin{subequations}}
\def\es{\end{subequations}}
\def\g{\gamma}
\def\G{\Gamma}
\def\vp{\varphi}
\def\mpl{M_{\rm P}}
\def\ms{M_{\rm s}}
\def\ls{l_{\rm s}}
\def\l{\lambda}
\def\gs{g_{\rm s}}
\def\d{\partial}
\def\co{{\cal O}}
\def\sp{\;\;\;,\;\;\;}
\def\spa{\;\;\;}
\def\r{\rho}
\def\dr{\dot r}
\def\dt{\dot\varphi}
\def\e{\epsilon}
\def\k{\kappa}
\def\m{\mu}
\def\n{\nu}
\def\om{\omega}
\def\tn{\tilde \nu}
\def\p{\phi}
\def\vp{\varphi}
\def\r{\rho}
\def\s{\sigma}
\def\t{\tau}
\def\a{\alpha}
\def\b{\beta}
\def\de{\delta}
\def\bra#1{\left\langle #1\right|}
\def\ket#1{\left| #1\right\rangle}
\newcommand{\stt}{\small\tt}
\renewcommand{\theequation}{\arabic{section}.\arabic{equation}}
\newcommand{\eq}[1]{equation~(\ref{#1})}
\newcommand{\eqs}[2]{equations~(\ref{#1}) and~(\ref{#2})}
\newcommand{\eqto}[2]{equations~(\ref{#1}) to~(\ref{#2})}
\newcommand{\fig}[1]{Fig.~(\ref{#1})}
\newcommand{\figs}[2]{Figs.~(\ref{#1}) and~(\ref{#2})}
\newcommand{\GeV}{\mbox{GeV}}
\def\SIZE{1.00} 

\baselineskip .8cm   
  
\section{Introduction}  
\label{intro}  
The goal of this paper is to deepen the connection between the
Brane Gas Cosmology (BGC) presented 
in~\cite{Alexander:2000xv,Brandenberger:2001kj} and realistic models
of particle physics derived from superstring theory.  
The BGC model employs the Brandenberger-Vafa mechanism 
of~\cite{Brandenberger:1989aj}, in an attempt to understand the 
origin of our large $(3+1)$-dimensional universe, while simultaneously resolving
the initial singularity problem of standard Big Bang cosmology.

Although we have been relatively successful in keeping with these ambitions,
the price we pay is that we have made little connection with 
realistic particle physics.  Part of this problem arises from the 
toroidal compactification of superstring theory used in~\cite{Alexander:2000xv}.
It is well known that compactifications of superstrings on manifolds of trivial
holonomy cannot
produce realistic models of particle physics.  

The original setting for the BGC model was within Type IIA string theory.
In this paper we present modifications of BGC by studying the model within the context of 
other branches of the $M$-theory moduli space.
We consider the physics of the resulting
$p$-brane gases on manifolds with non-trivial holonomy.  In particular, we focus
our attention on a manifold with a $K3$ subspace.  Such compactifications
are of interest for numerous reasons, as we shall see below. 
For one, it seems that the mathematics of $K3$ is intimately connected
to the heterotic string, which is the superstring theory that is most easily
related to realistic particle physics.

We use the fact that a certain string theory compactified on $K3 \times \mT^2$ can be related
via duality to another string theory on a Calabi-Yau three-fold 
with $SU(3)$ holonomy.  The general results 
of~\cite{Alexander:2000xv} remain intact.  Using the rich properties
of $K3$ surfaces and the dualities which (with strings) tie together the various 
branches of the $M$-theory moduli space, we come ever closer to the construction of a potentially 
realistic model of the universe.  

Of course, the only physically relevant case is that of an $\N=1$
theory in four dimensions.  Unfortunately, we do not produce such a theory here!
However, some of the models presented in this paper provide a significant improvement
to the toroidal compactifications of~\cite{Alexander:2000xv}.  We provide
an existence proof of a brane gas cosmology with an $\N=2$ theory in four 
dimensions compactified on a Calabi-Yau three-fold with $SU(3)$ holonomy.  
Despite the lack of one-cycles in such spaces there are cases where
only three spatial dimensions can become large.
It is our hope that such theories may provide clues to the behavior of more
realistic, $\N=1$ models.

The brane gas model solves the dimensionality problem, possibly revealing
the origin of our four-dimensional universe.  The model also solves the initial singularity
and horizon problem of the standard Big Bang cosmology without
relying on an inflationary phase. Note that we do not exclude the possibility
of an inflationary phase (perhaps along the lines of the string-inspired inflationary
models of~\cite{Burgess:2001fx}~-~\cite{Turok:1988pg})
during some stage in the evolution of the universe.

The presentation of this paper is as follows. Section~\ref{dimp} introduces
the dimensionality problem as a problem for both string theory and cosmology.
We discuss the fact that eleven-dimensional supergravity places an upper bound
on the number of spatial dimensions in our universe and introduce the model
of Brane Gas Cosmology in section~\ref{bgc} in an attempt to explain
why we live in a three-dimensional world.  Section~\ref{bgk3} introduces the 
first steps toward generalising the model of Brane Gas Cosmology to 
manifolds with nontrivial holonomy.  Various scenarios are constructed from
different branches of the $M$-theory moduli space and in different background
topologies.  We then attempt to relate the different constructions via
dualities.  We argue that these cosmological scenarios provide further
evidence for the conjectured dualities.  Some final thoughts are presented in section~\ref{conc}.
\subsection{The Dimensionality Problem}
\label{dimp}  
Arguably, one of the most significant dilemmas in string theory is the
dimensionality problem. A consistent formulation of superstring theory
requires the universe to be $(9+1)$-dimensional but 
empirical evidence demonstrates that the universe is $(3+1)$-dimensional.  

One resolution to this apparent conflict 
is to hypothesise that six of the spatial dimensions are curled 
up on a near Planckian sized manifold, and are therefore difficult to
detect in the low energy world that we live in.  
But if this is the case, the question naturally arises, why
is there a difference in size and structure between our large $3$-dimensional
universe and the $6$-dimensional compact space?  What physical
laws demand that spacetime be split in such a seemingly unusual way?  Since we
are assuming superstring theory is the correct theory to describe the
physical universe, the answer to these questions must come from within the
theory itself.

Although the dimensionality problem is a very severe problem from a 
cosmological viewpoint it is rarely addressed.  For example, 
brane world cosmological models derived from string theories 
typically impose the identification
of our universe with a $3$-brane.  All current formulations fail to explain why
our universe is a $d$-brane of spatial dimension $d=3$, opposed to any other
value of $d$, and furthermore fail to explain why our universe is this particular
$3$-brane opposed to any other $3$-brane which may appear in the theory.  Due
to the current unnatural construction of such models, it seems possible that they will
inevitably require some form of the anthropic principle in order to address the
dimensionality problem.

The dimensionality problem is not unique to string theory, however;
it is an equally challenging problem for cosmology.  A truly complete
cosmological model (if it is possible to obtain such a thing), whether derived from
$M$-theory, quantum gravity or any other theory, should necessarily
explain why we live in $(3+1)$-dimensions.

Because this conundrum is an integral part of both superstring theory and
cosmology, it seems likely that only an amalgamation of the two will
be capable of producing a satisfactory solution.  After all, if one is going to
evolve from a $9$-dimensional space to a $3$-dimensional space, one
is going to require dynamics, and the dynamics of our universe are 
governed by cosmology.
\subsection{Brane Gas Cosmology}
\label{bgc}  
The search for an explanation of spacetime structure and dimensionality
is not completely hopeless.  It is possible
to show that the largest dimension in which one can construct a supergravity
theory is in $D=11$, thus providing an upper limit on the number
of spacetime dimensions~\cite{Nahm:1978tg}.~\footnote{By a supergravity theory
we mean a theory containing particles with a maximum spin of 2.}
Eleven dimensional supergravity is of 
particular interest since it
has been identified as the low energy limit of 
$M$-theory~\cite{Townsend:1995kk,Witten:1995ex}.  
Because of this conjecture the theory provides an ideal starting point for the construction
of an $M$-cosmology.  

Eleven dimensional supergravity contains only three
fields: the vielbein $e^A_M$ (or equivalently the graviton $G_{MN}$), 
a Majorana gravitino $\psi_M$ and a three-form
potential $A_{MNP}$.  Here the indices are eleven-valued.  
In order to have equal numbers of bosonic and 
fermionic fields one must have the following number of degrees of freedom
for each of these fields: $e^A_M = 44$ components,  $\psi_M= 128$ components
and $A_{MNP}= 84$ components (see, for example~\cite{West:1998ey}).  The
full Lagrangian for this theory was first written down by Cremmer,
Julia and Scherk in~\cite{Cremmer:1978km} and will serve as our starting
point:
\begin{eqnarray}  
\L= 
&&\hspace{-6mm}  
- {1 \over 2\k^2}eR - \frac{1}{2} e\bar\psi_M \Gamma^{MNP}
D_N \left( \frac{1}{2} (\om + \hat\om) \right) \psi_P -\frac{1}{48}eF^2_{MNPQ}  
\nonumber \\  
&&\hspace{-6mm}  
-{\sqrt{2}\k \over 384}e   
\left(\bar\psi_M \Gamma^{MNPQRS}\psi_S + 12 \bar\psi^N \Gamma^{PQ}\psi^R \right)
(F + \hat F)_{NPQR}
\nonumber \\  
&&\hspace{-6mm}  
-{\sqrt{2}\k \over 3456}  
\epsilon^{M_1 \cdots M_{11}}F_{M_1 \cdots M_4}F_{M_5 \cdots M_8}A_{M_9 M_{10} M_{11}}
\,.
\label{11action}  
\end{eqnarray}    
Here $F_{MNPQ}$ is the curl (field strength) of $A_{MNP}$ 
and $\hat F_{MNPQ}$ is the supercovariant
$F_{MNPQ}$.  The spin connection is $\om_{MAB}$ and is given by the solution
to the field equation that results from varying it as an
independent field.  $\hat \om_{MAB}$ is the supercovariant connection given by
\be\label{scc}
\hat \om_{MAB} = \om_{MAB} + \frac{1}{8}\bar\psi^P\Gamma_{PMABQ}\psi^Q
\,.
\ee
The $\Gamma$'s in \eq{11action} are antisymmetrized products with unit weight,
\be
\Gamma^{AB} = \frac{1}{2}(\Gamma^{A}\Gamma^{B} -\Gamma^{B}\Gamma^{A}))
\,,
\ee
where
$\Gamma^{M} = e^M_A \Gamma^{A}$ are the Dirac matrices which obey 
$\{\Gamma^{A},\Gamma^{B}\} = 2 \eta_{AB}$.

The corresponding super-algebra in
terms of the central charges $Z$ is
\be\label{11al}
\{Q_\a,Q_\b\} = (\G^M C)_{\a \b} \, P_M + (\G^{MN}C)_{\a \b} Z_{MN}
	+  (\G^{MNPQR}C)_{\a \b} Z_{MNPQR}
\,,
\ee
where each central charge term on the right corresponds to a $p$-brane.
The $Q_\g$ are the supersymmetry
generators and the $C$ matrices are real antisymmetric 
matrices~\cite{Kaku:1999yd}.
This action will be the starting point
for the model of Brane Gas Cosmology presented below. 
As we have already 
mentioned, supergravity seems to place an upper
limit on the number of spacetime dimensions in our universe.
Other considerations which 
involve dynamical processes may in fact tell us why we are living
in three spatial dimensions.
In addition to eleven-dimensional $\N=1$ supergravity, the moduli space of 
$M$-theory also contains the five consistent superstring theories.
Because all of these theories are on equal footing (from a mathematical
perspective), it is likely that valuable information may be gained by
considering string cosmology in 
every branch of the $M$-theory moduli space.  We will explore
several such avenues and find interesting relations between them via the conjectured
web of dualities which link the various branches of the $M$-oduli space.
It is possible that cosmological considerations will provide clues as to 
why the $M$-theory uni-verse is fragmented into the superstring
penta-verse, leading to deeper insight into the nature of duality
symmetries.
 
Recently, a new way to incorporate $M$-theory into cosmology was introduced
in~\cite{Alexander:2000xv}.  
The motivations for this scenario, are the problems of the
standard Big Bang model (such as the presence of an initial singularity),
the problems of string theory (such as the dimensionality problem), and are cosmological.  
By cosmological we mean that we wish to stay in close contact with the standard Big Bang
model, and therefore maintain the initial conditions of a hot, small
and dense universe.  In our opinion, many other attempts to incorporate
$M$-theory into cosmology, such as the existing formulations of brane world scenarios, are 
often motivated
from particle physics and make little connection with what we know about
the origins of the universe.  They also suffer from the dimensionality
problem mentioned above.  Why do the extra dimensions have the topologies they do?
Why should a $3$-brane be favored over any other
$p$-brane for our universe, and why should we live on one particular $3$-three
brane versus another?  

Besides keeping close ties with our beloved Big Bang cosmology we want
to derive our model from the fundamental theory 
of everything, namely $M$-theory.  The difficulty here is that we don't know what
$M$-theory is.  We will therefore start with what we do know, the conjectured low energy
limit of $M$-theory, which is the eleven-dimensional, $\N =1$ supergravity
given by the action~(\ref{11action}).  Dimensional reduction of
this theory on $\mS^1$ results in Type IIA, $D=10$ supergravity which is the low energy
limit of Type IIA superstring theory.  
The overall spatial manifold was assumed to be toroidal in all nine spatial dimensions.  The bosonic part of
the low energy effective action for the background spacetime is
\begin{eqnarray} \label{bulk}
S_{NS} \, = \, {1 \over {2 \kappa^2}} \int d^{10}x \sqrt{-G} e^{-2 \Phi} 
\bigl[ R &+& 4 G^{\mu \nu} \nabla_\mu \Phi \nabla_\nu \Phi \nonumber \\
&-& {1 \over {12}} H_{\mu \nu \alpha}H^{\mu \nu \alpha} \bigr] \, ,
\end{eqnarray}
where we have included the Neveu-Schwarz - Neveu-Schwarz (NS-NS) fields only, and 
ignored the terms bilinear
in Ramond - Ramond (RR) fields.  Here $G$ is the determinant of the background metric 
$G_{\mu \nu}$, $\Phi$ is the dilaton, $H$ denotes the 
field strength corresponding to the bulk antisymmetric tensor 
field $B_{\mu \nu}$, and $\kappa$ is determined by the 10-dimensional 
Newton constant.

The supersymmetry algebra for the Type IIA theory is obtained by
dimensional reduction of the eleven-dimensional super-algebra of~\eq{11al} 
and is given by
\begin{eqnarray} \label{10al}
\{ Q_\a , Q_\b \}  \,=  \,(\G^M C)_{\a \b} \, P_M + (\G_{11}C)_{\a \b}\, Z 
	 \,+  \,(\G^{M}\G_{11}C)_{\a \b} \, Z_{M}
	 \,+ \, (\G^{MN}C)_{\a \b}\, Z_{MN} 		\nonumber	\\
	 \,+ \, (\G^{MNPQ}\G_{11}C)_{\a \b}\, Z_{MNPQ}
	 \,+ \,  (\G^{MNPQR}C)_{\a \b}\,\, Z_{MNPQR}
\,.
\end{eqnarray}
The universe is assumed to be
filled with a gas of the $p$-branes contained in the spectrum of 
this theory.  Furthermore, the torus is assumed to start out small (string length scale) and
with all fundamental degrees of freedom near thermal equilibrium.  
$M$-theory contains a $3$-form tensor
gauge field $B_{\m\n\r}$ which corresponds to an electrically charged
supermembrane (the $M2$-brane).  The $M5$-brane is a magnetically charged object.  
$M$-theory also contains the graviton 
(see, for example~\cite{Polchinski:1998rr}).  The $M2$-brane and $M5$-brane
can wrap around the $\mS^1$ in the compactification down to ten-dimensions,
and hence produce the fundamental string and the 
$D4$-brane of the Type IIA theory, respectively.  If the $M2$ and $M5$ branes
do not wrap on the $\mS^1$ they produce the $D2$ and $5$-brane solutions
in the IIA theory.  The graviton of $M$-theory obviously cannot wrap around the
$\mS^1$ and correspond to $D0$-branes of the IIA theory.  Finally, we have
the $D6$-brane, whose field strength is dual to that of the $D0$-brane
and the $D8$-brane which may be viewed as a source for the 
dilaton field~\cite{Polchinski:1998rr}.~\footnote{Non-string theorists
may be interested to note that the $p$-brane spectrum of a theory 
may simply be read off from the supersymmetry charges $Z_{\m_1} , \dots Z_{\m_p}$
present in the theory.
From the algebras (\ref{11al}) and (\ref{10al}) we can easily identify the
$p$-branes of eleven-dimensional supergravity and Type IIA superstring
theory respectively.  The dual $q$-branes of the $p$-branes
are found using the relation $p+q=D-4$.  For example,
in ten-dimensions the dual of the electrically charged $0$-brane is the 
magnetically charged $6$-brane.}

To summarise, the brane gas we are interested in contains $D$-branes of
even dimension $0,2,4,6$ and $8$, and odd dimensional $p$-branes with
$p = 1, 5$.  Therefore, the total action governing the dynamics of the
system is the background action~(\eq{bulk}) plus the action describing
the fluctuations of all the branes in the theory.

The action of an individual $p$-brane is the Dirac-Born-Infeld action~\cite{Polchinski:1998rq}
\begin{equation} \label{brane}
S_p \, = \, T_p \int d^{p + 1} \zeta e^{- \Phi} \sqrt{- det(g_{mn} + b_{mn} + 2 \pi \alpha' F_{mn})}
\end{equation}
where
\be\label{tension} 
T_p ={{\pi} \over {g_s}} (4 \pi^2 \alpha')^{-(p + 1)/2}
\,,
\ee
is the tension of the 
brane, $g_{mn}$ is the induced metric on the brane, $b_{mn}$ is the induced antisymmetric tensor field, and $F_{mn}$ is the field strength tensor of gauge fields $A_m$ living on the brane.~\footnote{From \eq{tension} we see that all of the branes in this scenario have {\it positive} tensions.} The constant $\alpha' \sim l_{st}^2$ 
is given by the string length scale $l_{st}$, and $g_s$ is 
the string coupling parameter.
The total action is the sum of the 
bulk action (\ref{bulk}) and the sum of all of the brane actions (\ref{brane}), each coupled as a delta function source (a delta function in the directions transverse to the brane) to the 10-dimensional action. 

The induced metric on the brane $g_{mn}$, with indices $m,n,...$ denoting space-time dimensions parallel to the brane, is determined by the background metric $G_{\mu \nu}$ and by scalar fields $\phi_i$ living on the brane (with indices $i,j,...$ denoting dimensions transverse to the brane) which describe the fluctuations of the brane in the transverse directions:
\begin{equation} \label{indmet}
g_{mn} \, = \, G_{mn} + G_{ij} \partial_m \phi_i \partial_n \phi_j + G_{in} \partial_m \phi_i \, .
\end{equation}
The induced antisymmetric tensor field is
\begin{equation} \label{indten}
b_{mn} \, = \, B_{mn} + B_{ij} \partial_m \phi_i \partial_n \phi_j + B_{i[n} \partial_{m]} \phi_i \, .
\end{equation}
In addition,
\begin{equation}
F_{mn} \, = \, \partial_{[m} A_{n]} \, .
\end{equation}

The evolution of the system described above, with Friedman-Robertson-Walker background
$G_{\mu \nu} \, = \, a(\eta)^2 diag(-1, 1, ..., 1)$, was analyzed in~\cite{Alexander:2000xv} and
discussed in greater detail in~\cite{Brandenberger:2001kj}.  The results 
were that the winding $p$-branes introduced a confining potential for the
scale factor $a(\eta)$ implying that these states tend to \it prevent \rm the universe
from expanding~\cite{Alexander:2000xv,Tseytlin:1992xk}.

A summary of the evolution of the universe according to the Brane Gas model
is the following.
The universe starts out small, hot and dense, toroidal in
all nine-spatial dimensions and filled with a gas of $p$-branes.  The $p$-branes
exhibit various behaviors.  They may wrap around the cycles of the torus
(winding modes), they can have a center-of-mass motion along the
cycles (momentum modes) or they may simply fluctuate in the bulk space 
(oscillatory modes).  By symmetry, we assume that there are equal numbers of winding and anti-winding modes.
When a winding mode and an anti-winding mode interact they unwind and form a loop in
the bulk spacetime.  As the universe tries to expand, the winding modes become 
heavy and halt the expansion.  Spatial dimensions can only
dynamically decompactify if the winding modes disappear.  

A simple counting
argument demonstrates that a $p$-brane winding mode and a $p$-brane 
anti-winding mode are likely to interact in at most $2p+1$ 
dimensions.~\footnote{For example, consider two
particles (0-branes) moving through a space of dimension $d$.  These particles
will definitely interact (assuming the space is periodic) if $d=1$, whereas they
probably will not find each other in a space with $d>1$.}
In $d = 9$ spatial dimensions, there are no obstacles preventing the disappearance 
of $p=8, 6, 5$ and $p=4$ winding modes, whereas the lower dimensional brane 
winding modes will allow a hierarchy of dimensions to become large. 

For volumes large compared to the string volume, the 
$p$-branes with the largest value of $p$ carry the most energy (see \eq{pbrane}), and therefore
they will have an important effect first. 
The 2-branes will unwind in $2(2)+1=5$ spatial dimensions allowing these dimensions to become large. 
Within this distinguished $\mT^5$, the 1-brane winding modes 
will only allow a $\mT^3$ subspace to become large.
Hence the above model provides a dynamical decompactification mechanism
which results in a macroscopic, $3$-dimensional universe,
potentially solving the
dimensionality problem discussed in section~\ref{dimp}.

We have reiterated the general arguments of~\cite{Alexander:2000xv}
in order to point out the special way in which the dimensional hierarchy
is made manifest.  This decomposition into products of spaces exhibiting these particular
dimensionalities will be of interest to us in the following
sections.  A careful counting of dimensions leads to the resulting manifold
\be\label{manif}
\mathbf{\M}^{10}_{IIA} = \mS^1 \times \mT^4 \times \mT^2 \times \mT^3
\,,
\ee
where the $\mS^1$ comes from the original compactification of $M$-theory
and the hierarchy of tori are generated by the self-annihilation of $p=2$ and $p=1$ branes
as described above. 

Note that the compact dimensions remain compact due to remnant winding states.
For example, the $\mT^2$ remains small due to string winding modes which cannot self-annihilate once the $\mT^3$ has grown large. (These remnant winding modes have coordinates
in the large $\mT^3$ as well.) However, one issue which remains a concern is the
stability of the radius of the compact dimensions to
inhomogeneities as a function of the three coordinates
$x_i$; $i = 1, 2, 3$ corresponding to the large spatial
dimensions. The separation in $x_i$ between the branes wrapping the 
small tori is increasing, and there appears to be no mechanism to keep the
``internal" dimensions from expanding inhomogeneously in $x_i$
\it between \rm the branes. A simple but unsatisfactory solution, is to
invoke a non-perturbative effect similar to what is needed to
stabilize the dilaton at late times, namely to postulate
a potential which will stabilize the moduli 
uniformly in $x_i$ (after the winding modes around the $x_i$
directions have disappeared). Work on a possible solution within
the context of the framework presented here is in progress.~\footnote{We 
thank R. Brandenberger and D. Lowe for discussions on this issue.}

From \eq{manif} it appears that the universe may have undergone
a phase during which physics was described by an effective six-dimensional
theory.  It is tempting to draw a relation between this
theory and the scenario of~\cite{Arkani-Hamed:1998rs,Antoniadis:1998ig}.
However, since the only scale in the theory is the string scale it
seems unlikely that the extra dimensions are large enough
to solve the hierarchy problem.  This will be studied in a future
publication.
\section{Brane gases on K3 and Calabi-Yau manifolds}
\label{bgk3}  
The greatest deficiency of the brane gas model is its current lack of contact with particle physics.  
Part of this problem is the fact that the compactification was carried out 
on a toroidal manifold which possesses trivial holonomy.  
Because of this it is impossible to reproduce
a realistic model of particle physics (consistent with string theory), and hence
the resulting cosmological model cannot be a realistic one.  
We will now try to generalise the model of brane gas cosmology
to $K3$ and, via duality, to Calabi-Yau three-folds.
\subsection{Brane gases and heterotic $E_8 \times E_8$ strings}
\label{bghet}  
The cosmological scenario of~\cite{Alexander:2000xv} was developed within
the context of $M$-theory on $\mS^1$ which gives Type IIA supersting theory
in ten-dimensions.  Let us begin by considering the same scenario
in a different branch of the $M$-theory moduli space, one which is 
more easily connected to realistic particle physics, the $E_8 \times E_8$
heterotic string theory.  Like the Type IIA string, the $E_8 \times E_8$
theory can be obtained directly from eleven-dimensional 
$M$-theory~\cite{Horava:1996qa}.  
The specific conjecture is that strongly coupled $E_8 \times E_8$ heterotic
superstring theory is equivalent to $M$-theory on the orbifold 
$\mS^1/\mZ_2 \times \mR^{10}$~\cite{Horava:1996qa}.

To maintain the initial conditions of~\cite{Alexander:2000xv}, we will
assume that the nine-spatial dimensions are periodically identified so that
the full ten-dimensional spatial manifold is
\be
\M_{het}^{10} = \mS^1/\mZ_2 \times \mT^9
\,,
\ee
where the $\mS^1/\mZ_2$ is from the initial compactification of
eleven-dimensional $M$-theory. The ten-dimensional low energy effective action 
is
\be
\label{hetaction}
S_{het}(g,A,\Phi,\dots) = \frac{1}{2\k^2} \int d^{10}x \sqrt{-g} \, e^{-2 \Phi}\, 
\bigl[\, R(g) + 4(\nabla \Phi)^2 - \frac{1}{3} H^2 + \frac{\a'}{30} T\!r \, F^2_A + \cdots \,\big]
\,,
\ee
for a metric $g$ of signature $(-,+,\cdots,+)$, a connection $A$ and the
dilaton $\Phi$.  The fields denoted by the ``$\cdots$'' have been
omitted due to their irrelevance to the current topic of interest.  They include the
supersymmetric partners of the fields listed,  for example; the dilatino,
the gravitino and the gaugino.   
 
Among the fundamental constituents of the heterotic theory
are 2-dimensional objects (from the $M2$-brane in the eleven-dimensional
theory) and strings.  As described in section~\ref{bgc}, these $p =2$ and $p=1$ dimensional
objects will be
the only winding modes which are important for the decompactification
of a nine-dimensional space.  The hierarchy in the sizes of the growing dimensions,
will therefore be the same as that presented in section~\ref{bgc} formulated
within the context of Type IIA string theory (see~\eq{manif}).  
The entire ten-dimensional manifold decomposes into
\be\label{manhet}
\mathbf{\M}^{10}_{het} = \mS^1/\mZ_2 \times \mT^4 \times \mT^2 \times \mT^3
\,.
\ee
Note, that \it before \rm the string winding modes have annihilated the space will
look like $\M^{10} = \mS^1/\mZ_2 \times \mT^4 \times \mT^5$ which may be
approximated by the heterotic $E_8 \times E_8$ theory compactified on $\mT^4$.  
Let us take a moment to examine this theory in more detail.
\subsubsection{\it Heterotic $E_8 \times E_8$ on $\mT^4$}
\label{hett4}  
The six-dimensional heterotic theory, achieved by dimensional reduction
of the action (\ref{hetaction}), is
\be
\label{het6}
S_{\mT^4}^{het} = \frac{1}{2\k_6^2} \int d^{6}x \sqrt{-g_6} \, e^{-2 \Phi_6}\, 
\bigl[\, R + 4(\nabla \Phi)^2 - \frac{1}{2} |\tilde H_3|^2 - 
\frac{\k^2_6}{2g^2_6}\,|F_2^2| \,\big]
\,,
\ee
where the six-dimensional 
dilaton is dependent on the internal space volume and 
$\Phi$:~\footnote{Here we use the notation of~\cite{Polchinski:1998rr}.}
\be
e^{-2 \Phi_6} = V e^{-2 \Phi}
\,.
\ee
The resulting theory is an $\N = 2$, non-chiral theory.

The decomposition of~\eq{manhet} is nearly identical to the 
decomposition of $\mathbf{\M}^{10}_{IIA}$ in
\eq{manif}.  Considering these similarities, one is tempted to suspect a relation
between the two superstring theories, Type IIA and heterotic $E_8 \times E_8$.
Of course, such a duality exists and is well 
known~\cite{Witten:1995ex,Hull:1995ys}.  Let us return to this issue
after briefly introducing the surface $K3$ which will play a key role is our
discussion of duality.

\subsubsection{$K3$ surfaces}
There are only two Ricci-flat, K\"{a}hler manifolds in four-dimensions.
One we have already discussed is the torus $\mT^4$ and the other is
the surface $K3$.~\footnote{By ``surface'' we mean two complex dimensions.}  
The surface $K3$ is a Calabi-Yau manifold but it
has only two complex dimensions ($d=4$ real dimensions) and $SU(2)$
holonomy.  This surface has appeared extensively in the literature
and plays a paramount role in the analysis of string duality symmetries.
An extensive introduction to the properties of $K3$ can be found 
in~\cite{Aspinwall:1996mn}.  

Although no explicit construction of the metric on $K3$ has been
found~\cite{Yau:1978}, it is possible to construct the manifold
from an orbifold of $\mT^4$~\cite{Page:1978zu}.  One starts by
identifying coordinates of $\mT^4$, $x^i \sim x^i + 2\pi$ and
then making the identification $x^i \sim -x^i$, where $x^i \in \mR^4$.
This will lead to $16$ fixed points located at $x^i = \pi n$, where
$n \in \mZ$.  By removing the points as four-spheres and then 
filling the $16$ resulting holes with four-dimensional Eguchi-Hanson instantons,
it is possible to achieve an approximation to the smooth Ricci-flat
$K3$ manifold.~\footnote{Note that any two $K3$'s are diffeomorphic to each other, and therefore
we can produce all of the topological invariants of $K3$ from just one example.}
This approximation becomes more and more precise in the
limit as the radius of the instantons approaches zero~\cite{Lu:1999xt}.
For the purposes of this paper we may consider $K3$ as 
\be\label{k3t}
K3 \simeq \mT^4/\mZ_2
\,.
\ee
\subsubsection{\it Heterotic $E_8 \times E_8$ and Type IIA}
\label{e82a}  
Now consider ``moving'' the $\mZ_2$ symmetry in \eq{manhet} from the $\mS^1$ to
the $\mT^4$.  We then have the structure
\be\label{IIAk3}
\mathbf{\M}^{10}_{} = \mS^1 \times \mT^4/\mZ_2 \times \mT^2 \times \mT^3
\,.
\ee
From the full, eleven-dimensional perspective
the \eq{IIAk3} looks like $M$-theory compactified on $\mS^1 \times K3$ (recall~\eq{k3t}), which 
is Type IIA string theory compactified on $K3$.  In fact, this \it gives \rm the duality between
this theory and the one presented in section~\ref{hett4}. 
 
Type IIA superstring theory compactified on $K3$ is
dual to heterotic $E_8 \times E_8$ superstring theory compactified on
$\mT^4$~\cite{Witten:1995ex,Hull:1995ys}.  The duality seems almost obvious
when one considers the manifolds (\ref{IIAk3}) and (\ref{manhet}),
and recalls the ability of $K3$ compactifications to break supersymmetry.
Because the heterotic theory has less supersymmetry then the IIA theory, some of the
supersymmetries of the IIA theory must be broken before any identification
between the two theories can be made.  A $\mT^4$ compactification of the
IIA theory is a $\N=4$ theory. As discussed above, the properties of $K3$
reduce the supersymmetry of the compactified IIA theory down to $\N = 2$, which
is the number of supersymmetries in the theory obtained by compactification
of the heterotic string on $\mT^4$.

These cosmological considerations seem to provide further evidence for the
conjectured duality.
To make this equivalence more transparent let us consider the theory obtained
from compactification of Type IIA on $K3$, and then compare this to our
analysis of the heterotic string on $\mT^4$ (section~\ref{hett4}).
\subsection{Type IIA on $K3$}
\label{IIAonk3}  
The simplest compactification to four-dimensions other than the torus
is the surface $K3$ (having $SU(2)$ holonomy) along with a cartesian product
of $\mT^2$.  We have mentioned that $K3$ is a very special surface
within the context of string theory.  In the next few sections we continue to
investigate the interesting physics that arises in $K3$ compactifications
and consider some of the implications this surface has on string cosmology.

Our first obstacle to generalising the brane gas model of cosmology
to $K3$ and Calabi-Yau spaces 
is the absence of one-cycles on both of these objects.~\footnote{For
$K3$ surfaces and phenomenologically interesting Calabi-Yau spaces the first Betti 
number vanishes, $b_1 =0$ (see e.g. \cite{Aspinwall:1996mn,Kaku:1999yd}).}
At first glance the arguments of~\cite{Alexander:2000xv} seem to rely on the presence of
one-cycles in the compactified dimensions.  We will show that a more
careful consideration demonstrates that this is not necessarily the case. 

The reduction of Type IIA theory on $K3$, which is suggested by the
\eq{IIAk3}, leads to the following action
\be
\label{IIAk3ac}
S^{IIA}_{K3} = \frac{1}{2\k_6^2} \int d^{6}x \sqrt{-g_6} \, e^{-2 \Phi_6}\, 
\bigl[\, R + 4(\nabla \Phi)^2 - \frac{1}{2} e^{2 \Phi_6}|\tilde H_3|^2 - 
\frac{\k^2_6}{2g^2_6}\,|F_2^2| \,\big]
\,.
\ee
Notice that this action can be transformed into (\ref{het6}) by
the conformal transformation
\be
g_{\m\n} \rightarrow e^{2 \Phi_6} g_{\m\n}
\,,
\ee
along with the reflection 
\be\label{refl}
\Phi_6 \rightarrow - \Phi_6
\,.
\ee
Here, $\tilde H_3 \rightarrow e^{2 \Phi_6} *_6 \tilde H_3$,
where  $ *_6 $ results from the factorization of the ten-dimensional
$*$ by $*_{10} = *_6 *_4$~\cite{Polchinski:1998rr}.  Because of the 
identification of $\Phi_6 \rightarrow - \Phi_6$, the conjectured
duality maps a strongly coupled theory to a weakly coupled theory.

The IIA theory on $K3$ contains the IIA string formed by the wrapping
of $M2$-branes on the $\mS^1$ from the compactification of $M$-theory
(as described in section~\ref{bgc}).  There is another string in this theory
however, obtained by wrapping an $M5$-branes around the entire four-dimensional
$K3$.  This is exactly the heterotic string~\cite{Cherkis:1997bx}.  

The two theories (\ref{IIAk3ac}) and (\ref{het6}) have the same
low energy supergravity description~\cite{Hull:1995ys}, which is an $\N =2$ 
supergravity coupled to abelian super-Yang-Mills multiplets.  This results
in $80$ scalar fields which span the moduli 
space of vacua.~\footnote{At first glance, one may be concerned that the Type IIA theory lacks the $E_8 \times E_8$
Yang-Mills fields of the heterotic string.  This is not a problem however,
since the moduli space of the IIA theory is enhanced via the 
compactification onto $K3$, and the resulting theory has the expected $80$
scalar fields mentioned above~\cite{Kaku:1999yd,Aspinwall:1996mn}.}  
\subsubsection{\it $IIA$ brane gas on $K3$}
\label{iiabg}  
Let us consider a simple modification to the brane gas model 
summarised in section~\ref{bgc}.  The background will be $M$-theory on the manifold
\be\label{mfold}
\M = \mS^1 \times K3 \times \mT^5
\,,
\ee
or Type IIA string theory in a toroidal universe, compactified on $K3$.
The gas of branes are the branes of the Type IIA theory.  

Recall that string winding modes will act like rubber bands wrapped 
around the cycles of the toroidal universe and hence, their existence
tends to prevent the universe from expanding.  Branes of larger dimension
(greater values of $p$) have a similar behavior as we demonstrate below, but
the energies in these branes is greater than the energy in the
string winding modes, and therefore they have an important effect on the dynamics
of the universe first.  We will show that such winding branes may wrap around
cycles of $K3$ and prevent it from expanding. 

The energy of a $p$-brane winding mode in an FRW universe with scale factor $a(\eta)$  
can be calculated from the 
action~(\ref{brane}), and is given by
\be\label{pbrane}
E_p(a) \simeq T_p \, a(\eta)^p
\,,
\ee
where $\eta$ is conformal time 
and $T_p$ is the tension of the $p$-brane given by \eq{tension}.  From this
we see that the energy
of a $p$-brane winding mode increases with $p$.  The ratio of the energy in a $(p+1)$-brane winding mode to a $p$-brane winding mode in this background is
\be\label{pbraneen}
\frac{E_{(p+1)}}{E_p} \simeq (4\pi^2\alpha')^{-1/2} \, a(\eta)
\,.
\ee
Note that there is no $p$ dependence in this equation.
Equation~(\ref{pbrane}) is the energy of a $p$-brane
wrapped around a one-cycle.  This is the correct expression to
use when calculating the energy in winding modes wrapped around the toroidal
pieces of the manifold $\M$ in \eq{mfold}.  

We have mentioned above that $K3$ does not posses one-cycles and therefore
strings cannot wrap around this portion of the manifold and prevent it
from expanding.
However, $K3$ does have Betti number $b_2 = 22$ (twenty-two harmonic two-forms) and therefore contains $22$ two-cycles, which $p > 1$ branes can wrap around. Note that the torus $\mT^5$ also contains 
($b_2 = 10$) two cycles. 
 
For our cosmological considerations we take the initial state of the universe 
to be the same as that described
in section~\ref{bgc}, except for the topology of the space which is
now given by (in the $D=10$ dimensional string description)
$K3 \times \mT^5$.  Recall that the highest dimensional brane winding
modes have an effect on the dynamics first, since they have the largest energy
and therefore fall out of equilibrium first.
As described in section~\ref{bgc}, brane winding modes for $p=8,6,5,4$ branes
do not have an effect on the decompactification process.  
The first branes to have an effect are the $2$-brane winding modes.  
However, we shall assume that the $2$-branes
which are wrapped around the two-cycles of the $\mT^5$ are heavier
than the $2$-branes which are wrapped around the two-cycles of the $K3$;
hence, they will fall out of thermal equilibrium and consequently effect 
the dynamics of the decompactification first. A plausibility argument
for this assumption goes as follows: recall that the
energy of a membrane scales with its area.  Now let us consider the topological
structure of the two-cycles of $K3$ and $\mT^5$.  A theorem familiar to many 
mathematicians states that any compact connected surface with boundary
is toplogically equivalent to a sphere ($\mS^2$), the connected sum of $n$ tori 
($\mT^2 \# \, \cdots \# \, \mT^2$) , or the
connected sum of $n$ projective planes ($\mR\mP^2 \,\# \cdots \# \, \mR\mP^2$), 
with a finite number of discs removed.~\footnote{For readers not familiar with the
term, the \it connected sum \rm of the surfaces $S_1$ and $S_2$ is defined by
removing a small disc from each of $S_1$ and $S_2$, and gluing the boundary
circles of these discs together to form a new surface $S_1 \# S_2$.}  Therefore, the two-cycles under consideration 
must have one of the above mentioned topological structures.  Since neither 
$K3$ nor $\mT^5$ contain a m\"{o}bius band, we may rule out any topological structure containing the real projective plane $\mR\mP^2$ which is non-orientable.

Let us first consider the topology of the two-cycles in $\mT^5$.  The easiest
way to determine their topological structure is to consider the elementary case of $\mT^2$.  In the torus,
there is only $b_2(\mT^2) = 1$ two-cycle. Clearly, this two-cycle is just the entire $\mT^2$.  
Note that this two-cycle is made up of $b_1(\mT^2) = 2$ one-cycles.  From this we can 
easily generalize to the case of $\mT^5$, by considering the number of ways
the $\mS^1$'s in $\mT^5$ can hook up to form two-cycles.  There are five $\mS^1$'s in $\mT^5$
and hence ${}_5C_2 = 10$ \it toroidal \rm two-cycles in the space.  Because this agrees with 
the calculation of the Betti number $b_2(\mT^5) = 10$, we know we have taken into
account all of the two-cycles.

Now consider the two-cycles in $K3$.  Since $K3$ has no one-cycles ($b_1(K3)=0$)
and hence no $\mS^1$'s, 
it is impossible for the two-cycles in $K3$ to have $\mT^2 = \mS^1 \times \mS^1$ topology.  
Hence the $b_2 = 22$ two-cycles in $K3$ must be topologicaly equivalent to \it spheres \rm.
Because the two-cycles of $K3$ and $\mT^5$ differ in their topological structure
membranes will have to wrap differently on the distinct types of two-cycles of 
the product manifold $K3 \times \mT^5$.

Recall in the inital conditions of our  cosmological model we assumed all 
radii $r_i$ start out 
on equal footing, near the string length scale $r_i \approx R_{st}$, where 
$i = 1,\dots,9$.  A membrane wrapped around a spherical two-cycle in the $K3$ 
subspace will have energy propotional to its surface area $\mathcal{E}_{S^2} \propto 4 \pi R_{st}^2$.
The same membrane wrapped around a toroidal two-cycle of the $\mT^5$ would have
energy $\mathcal{E}_{T^2} \propto 4 \pi^2 R_{st}^2$.  Hence, the two-branes wrapped around
two-cycles of the $\mT^5$ will (on average) be heavier then the two-branes wrapped around
two-cycles of the $K3$ by a factor of $\pi$.

In short, for a $2$-brane in the space $K3 \times \mT^5$ it is energetically more
favorable to wrap around one of the $22$ two-cycles of the $K3$ then to wrap around one
of the $10$ two-cycles of
the $\mT^5$ (when all radii are equal). We come to the conclusion that it is probable that the wrapped $2$-branes
on $\mT^5$ will fall out of thermal equilibrium before the wrapped $2$-branes
on $K3$ and allow the $\mT^5$ to grow.  This argument is only a
plausibility argument due to our assumptions about initial conditions and setting all
radii to be near the string scale.  However, we believe that such initial
conditions are well motivated given the cosmological setting of~\cite{Alexander:2000xv}.

One possibility then (arguably the most probable and certainly the most 
desirable from the viewpoint of a realistic model)
is that the interaction of winding and antiwinding
$2$-branes causes the $\mT^5$ subspace to grow (since $2(2) + 1 =5$).~\footnote{
In order to get a very large gauge group, e.g. $E_8 \times E_8 \times SU(2)^4
\times U(1)^4$ it is probably necessary for the $K3$ space to be very small
(with volume of order one in units of $(\alpha')^2$) and singular~\cite{Aspinwall:1996mn}.}  
After that, there is no way for the $2$-branes wrapping the $K3$ space
to self-annihilate (by the old dimension counting argument) since these
branes have coordinates (fluctuations) in the $\mT^5$ space as well as in the $K3$.
This compact space remains small due to these remnant winding 
states.~\footnote{Note that similar arguments should apply to Calabi-Yau threefolds.}
Within the $\mT^5$ the string winding mode self-annihilation causes a 
$\mT^3$ torus to grow large.  The final result (nearly the same
as in the $\mT^9$ compactification of section~\ref{bgc}) is,
\be
\M^{9} = K3 \times \mT^2 \times \mT^3
\,.
\ee
The winding modes around the $\mT^3$ completely vanish~\cite{Brandenberger:2001kj}
and hence the $\mT^3$ topology now looks like $\mR^3$.
The $\mT^2$ remains small due to remnant string winding modes which cannot self-annihilate
once the $\mT^3$ has grown large. (Note that these winding modes have coordinates
in the large $\mT^3$ as well.)  
Because there is only one scale in the theory (the string scale) it
seems unlikely that the $\mT^2$ subspace will be much larger than
the $K3$ space.  In fact, the remnant winding states may cause the
$\mT^2$ to shrink back down to the string scale.  This implies that the 
four-dimensional theory is Type IIA string theory compactified
on $K3 \times \mT^2$.~\footnote{A full treatment of the IIA string theory compactified
on $K3 \times \mT^2$ is given in~\cite{Kiritsis:1998hj}.  A detailed analysis of the six-dimensional
theory which describes this shrinking $\mT^5$ fracturing into an expanding $\mT^3$
and shrinking $\mT^2$ is underway.  It appears that the $\mT^3$ space makes a graceful
exit from a contracting phase into an expanding phase, perhaps along the lines 
of~\cite{DeRisi:2001ed}.}

In this theory, branes with 
$p<5$ can wrap around the two-cycles of the $K3$, producing monopole
winding states on $\mT^2$.  The final four-dimensional
field theory has $\N=4$ supersymmetry and is identical to the low-energy field
theory description of the heterotic string on $\mT^6$~\cite{Hull:1995ys}.

The $K3$ compactification of IIA string theory presented above is an improvement
over the trivial toroidal compactification of~\cite{Alexander:2000xv},
in the sense that $K3$ has holonomy $SU(2)$ and we have reduced
the number of supersymmetries by half (from $\N=8$ to $\N=4$).  Of course, the only physically relevant
case is that of $D=4$, $\N = 1$.  Alas, we do not construct
such a theory
in this paper but we will produce a $D=4$, $\N = 2$ theory
in what follows.
\subsubsection{\it $E_8 \times E_8$ brane gas on $K3$}
\label{e8bg}  
One way to get an $\N = 2$ theory in four-dimensions is
to compactify the heterotic $E_8 \times E_8$ string on a complex three-fold
with $SU(2)$ holonomy.  All such manifolds are of the form
$K3 \times \mT^2$ (and its quotients)~\cite{Aspinwall:1996mn}.  Once again, we find that 
a brane gas within the context of heterotic string theory on a nine-dimensional
manifold $K3 \times \mT^5$, is governed by the same arguments presented
in section~\ref{bgc} (also see section~\ref{e82a}).  The symmetry group decomposes
as 
\be
SO(1,9) \supset SO(1,5) + SO(4) \simeq SO(1,5) + SU(2) + SU(2)
\,,
\ee
and we have an $\N=1$ theory in six-dimensions.
The $2$-branes 
allow the $\mT^5$ submanifold to grow and then within the $\mT^5$ 
the strings allow a $\mT^3$ space
to grow large.  From the $D = 11$ $M$-theory point of view,
the full $d=10$ manifold is
\be
\mathbf{\M}^{10}_{\mbox{\it \scriptsize het on}\, K3} = 
\mS^1/\mZ_2 \times K3 \times \mT^2 \times \mT^3
\,,
\ee
which gives an $\N = 2$ theory in $D=4$.
\subsection{Brane gases on Calabi-Yau manifolds}
\label{bgcy}  
A realistic construction of our universe based on superstring theory requires compactification onto spaces
of nontrivial topology.  If one chooses to compactify on a manifold, then
particle physics requires that this be a Calabi-Yau threefold $\mX$.~\footnote{If one is willing to allow the compactified space to have 
singularities then it is possible to generate realistic particle physics models
from orbifold compactifications.}

The first difficulty we encounter when considering Brane Gas Cosmology
and Calabi-Yau compactifications with $SU(3)$ holonomy is the absence of one-cycles in $\mX$
($b_1(\mX) =0$).  In general, such manifolds
do have higher dimensional $p$-cycles, and we may employ the
same arguments given in our discussion of $K3$ to show that the
absence of one-cycles may not pose a problem for the cosmological model 
(see section~\ref{iiabg}).

In fact, some properties of Calabi-Yau three-folds are easier to understand
within our context then the $K3$ surface.
In a Calabi-Yau three-fold $\mX$ the K\"{a}hler moduli parameterize the relative
resizing of the two-cycles (and the four-dimensional subspaces representing
their dual four-cycles).  Thus the K\"{a}hler moduli give the overall
scale of the manifold; i.e. fluctuations of these moduli correspond to
fluctuations in the volume of the space.  Deformations of the complex
structure moduli correspond to changing the ``shape'' of the manifold~\cite{Greene:1996cy}.

For $d=3$ complex manifolds with exactly $SU(3)$ holonomy, 
the fermion zero modes (which are
harmonic $(0,n)$ forms) must be covariantly constant. This implies
vanishing of the Hodge numbers $h^{2,0} = h^{0,2}=0$.  If this condition
is satisfied then the entire moduli structure of $\mX$ becomes a local product of
the complex structure moduli and the K\"{a}hler moduli.  Deformations in
these two types of structure moduli may then be studied independently.  In the
complex two-fold $K3$, $h^{2,0}=1$.  The complex and  K\"{a}hler moduli are
degenerate and deformations of these structures must be 
studied simultaneously~\cite{Aspinwall:1994rg}.

The product structure of the moduli on $\mX$ may simplify our study
considerably.  For example, if we choose $\mX$ with only one
 K\"{a}hler class then all even dimensional submanifolds in
$\mX$ will scale with a \it single \rm parameter (or the appropriate power thereof).
We may imagine then the possibility of constraining 
only one two-cycle
(perhaps with a winding brane) and thereby ``fix'' all other two-cycles
in $\mX$.~\footnote{We thank B.~Greene for pointing this out. After the completion 
of this work a related paper appeared in~\cite{Easther:2002mi}.}

The second difficulty (which will prevent us from achieving a realistic
$N=1$, $D=4$ cosmological theory) is that it is not clear how the manifold $\M^9$,
resulting from a direct compactification on $\mX$, will decompose.
Within the context of the brane gas model, all the manifolds $\M_i$ decompose
into products of a space and a five-dimensional torus due to the $2$-brane
winding modes (see, for example \eq{mfold}).  
In the case of a direct compactification onto $\mX$,
the full ten-dimensional manifold is
\be
\M = \mM^1 \times {\mX} \times \mT^3
\,,
\ee
where $\mM^1$ is either $\mS^1$ (for the Type IIA theory) or
$\mS^1/\mZ_2 $ (for the $E_8 \times E_8$ heterotic string).  It is not clear
what the five-dimensional subspace (analogous to the $\mT^5$ mentioned above) will be.

Despite this dilemma, there is good reason to believe that such a compactification
should somehow be compatible with the brane gas picture.  The motivation
for this belief results from our lessons about dual $\N=4$ theories presented 
in section~\ref{iiabg}. 

\subsubsection{\it $IIA$ brane gas on a Calabi-Yau threefold}
\label{iiacy}  

In addition to the discussion of section~\ref{e8bg}, a second way to construct an $\N =2$ theory in $D=4$ is to consider
a Type II theory compactified on a Calabi Yau three-fold $\mX$, which is a complex manifold with holonomy $SU(3)$.  In this section we will
provide an existence proof of the compatibility of Calabi-Yau
manifolds and the model of Brane Gas Cosmology.  

We would like to find a duality between the $\N=2$ theories described
here, and in section~\ref{e8bg}.  Such a duality exists~\cite{Kachru:1995wm,Ferrara:1995yx}:
for a certain type of Calabi-Yau three-fold ${\mX}$ with $SU(3)$ holonomy, 
Type II theories compactified on ${\mX}$ are dual to heterotic
$E_8 \times E_8$ on $K3 \times \mT^2$.

We will not investigate this duality in detail and the interested reader
should see, for example~\cite{Polchinski:1998rr,Aspinwall:1996mn}.  
Many Calabi-Yau manifolds are $K3$ \it fibrations\rm, i.e. locally
a product of $K3$ with a two-dimensional manifold.  That is, the
manifold $\mX$ is a fibration where the generic fiber is a $K3$ surface.

In general, the conformal field theories (CFTs)
arising from compactifications on Calabi-Yau manifolds have very different
geometric descriptions in terms of the K\"{a}hler and complex structure moduli.
This brings us to a short discussion of \it mirror 
manifolds \rm~\cite{Greene:1996cy}.~\footnote{Much of our discussion on
this subject will ``reflect'' Polchinski's book~\cite{Polchinski:1998rr}.}
It appears that certain Calabi-Yau manifolds come in mirror pairs,
$\M$ and $\W$ where the conformal field theory descriptions of the two
manifolds are isomorphic (related by $H \rightarrow -H$) but certain
geometric properties are reversed.  A geometric correspondence
between two such manifolds is known for at least one type of such mirror pairs.
These are related to so called Gepner models, for which a relation
between mirrors is
\be
\W = \M/\Gamma
\,,
\ee
where $\Gamma$ is some subgroup of the global symmetry group that commutes
with the spacetime supersymmetry group.  This ``twisting'' by 
$\Gamma$ results in orbifold singularities in the mirror space $\W$.
Note that in the above compactification of $E_8 \times E_8$ on $K3$,
the $K3$ surface is itself a fibration with generic fiber
given by a $\mT^2$.
Because of this duality between IIA on a Calabi-Yau manifold and the 
heterotic string on $K3 \times \mT^2$, we suspect that BGC is compatible
with Calabi-Yau compactifications.  Specifically, the brane gas cosmology of 
section~\ref{e8bg} should be dual to a brane gas cosmology on a Calabi-Yau
three-fold.  
\subsection{Type IIB brane gases}
\label{iibbg}  
The Type IIB theory is a theory with chiral $\N = 2$ supersymmetry.  It 
is not clear how to derive this theory from an eleven-dimensional supergravity
theory and therefore its connection to $M$-theory is not as transparent as
that of the Type IIA and heterotic theories.  Nevertheless, the Type IIB
theory occupies a space in the $M$-theory moduli space, and therefore we
should consider BGC within the context of the IIB theory.

For simplicity and to make a comparison with~\cite{Alexander:2000xv}
we consider the IIB brane gas in a toroidal background $\mT^9$.
However, our general statements concerning IIA on $K3$ also apply to 
the IIB theory.
The algebra for the Type IIB superstring, which has two chiral spinors
$Q^i_\a$ is
\begin{eqnarray} \label{10alb}
\{ Q^i_\a , Q^i_\b \}  \,=  \, \delta^{ij}(\cP \G^M C)_{\a \b} \, P_M 
	+ (\cP \G^{M}C)_{\a \b}\, \tilde Z^{ij}_M 
	 \,+  \,\epsilon^{ij}(\cP\G^{MNP}C)_{\a \b} \, Z_{MNP} \nonumber	\\
	 \,+ \, \delta^{ij}(\cP\G^{MNPQR}C)_{\a \b}\, Z^+_{MNPQR}
	 \,+ \, (\cP\G^{MNPQR}C)_{\a \b}\,\tilde Z^{+ij}_{MNPQR}
\,,
\end{eqnarray}
where $\cP$ is a chiral projection operator and the tilde refers to
traceless $SO(2)$ tensors~\cite{Kaku:1999yd}.
From the algebra we see that this theory contains odd-dimensional BPS states which
are $Dp$-branes with $p = 1,3,5,\dots$.  Following our usual arguments,
we find that the $p=1$ and $p=3$ winding branes will have an effect
on the decompactification dynamics of the nine-dimensional universe.  The $3$-branes
allow a $\mT^7$ subspace to become large and then 
the string winding modes (as usual) result in a $\mT^3$ large sub-subspace.  
The overall nine-dimensional manifold $\M^9$ evolves
into
\be\label{miib}
\M^9_{IIB} = \mT^2 \times \mT^4 \times \mT^3
\,.
\ee

By comparing \eq{miib} with the corresponding decomposition within the
context of the IIA theory (\ref{manif}), we see the same overall structure
except for the switching of the roles of $\mT^2$ and $\mT^4$ within the scaled hierarchy of
large dimensions.  

Note, that the IIA string theory compactified on a circle of radius $R$, is dual to
Type IIB string theory compactified on a circle of radius $1/R$ (at the
same value of the coupling constant). In some sense, we may actually \it see \rm the effects of $T$-duality by comparing
\eq{manif} and \eq{miib}.  If we identify one of the $\mS^1$'s in the
$\mT^2$ as the $\mS^1$ that is transformed by $T$-duality, it makes sense that
the $\mT^2$ of one theory will be of smaller area than the $\mT^2$ of the other theory.
Of course $T$-duality alone does not explain why the area of the $\mT^2$ in the IIB theory
should be smaller then that of the $\mT^2$ in the IIA theory.  This difference
is determined by the brane spectrums of the IIA and IIB gases and the dynamics
arising from their respective cosmologies.

The $T$-duality relation between the Type IIA and Type IIB theories 
provides an interesting context in which to
explore mirror symmetry~\cite{Polchinski:1998rr,Strominger:1996it}.  Consider the
IIA string on a Calabi-Yau manifold $\M$ (section~\ref{iiacy}).  The manifold of states
of a $D0$-brane make up $\M$ itself, since the $D0$-brane can live anywhere.
Now consider the dual IIB theory on the mirror manifold $\W$.  The $Dp$-branes
of the IIB theory can wrap around the non-trivial cycles of $\W$.  As we have
explained, these will be odd $p$, $p$-branes and the Betti numbers of
$\W$ have $b_1 = b_5 =0$, 
which implies we must have $p=3$ winding branes.  As explained in~\cite{Polchinski:1998rr,Strominger:1996it}
this suggests a $T$-duality on three axes.  The $D0$-brane will have
three coordinates that map to internal Wilson lines
on the $D3$-brane, which must therefore be topologically a $\mT^3$.  Hence
$\W$ is a $\mT^3$ fibration and the mirror transformation is $T$-duality on
the three axes of $\mT^3$.  This implies that $\M$ must also be a 
$\mT^3$ fibration.  
\section{Conclusions}
\label{conc}  
The Brane Gas model of the early universe
provides a potential solution to the
dimensionality problem, which is a problem of both string theory and cosmology.  
Previous formulations of the model have failed to incorporate superstring 
compactifications capable of leading to realistic models of particle physics.  
We have taken the first steps toward modifying the scenario to accommodate compactifications 
on spaces with non-trivial holonomy.

Brane gases constructed from various branches of
the $M$-theory moduli space were analyzed.  By considering the dynamics of these gases 
in backgrounds of different topologies, we come to the conclusion that the general 
properties and successes of the cosmological model introduced in~\cite{Alexander:2000xv} 
remain intact.  In particular, we discussed compactification on manifolds with
non-trivial $SU(2)$ and $SU(3)$ holonomy, which correspond to $K3$ and
Calabi-Yau three-folds, respectively.  Despite the
lack of one-cycles around each dimension in these spaces, specific cases exist in which
only a three-dimensional subspace can become large (e.g., section~\ref{iiabg}).

Superstring duality symmetries become more lucid 
within this cosmological context.  Several examples are given where
brane gas models constructed from one sector of the $M$-theory moduli space
are linked to dual models in another.  We believe these considerations provide further
evidence for the conjectured duality symmetries of string theory. 

Finally, we constructed a model of cosmological brane gases from $E_8 \times E_8$ 
heterotic string theory on $K3 \times \mT^2$.  This gives an $\N=2$
theory in four-dimensions, on a manifold with nontrivial homology.  Such
an example is a significant improvement, from the point of view of
particle phenomenology, to the BGC of~\cite{Alexander:2000xv}.  Using the
power of duality we relate this theory to a brane gas model constructed
from Type IIA string theory compactified on a Calabi-Yau three-fold with
nontrivial $SU(3)$ holonomy.  We argue that this example provides an existence proof
of the compatibility between Brane Gas Cosmology and Calabi-Yau 
compactifications.

\section*{Acknowledgments}  
I would like to thank R.~Brandenberger for his comments on a draft of
this article and for numerous useful discussions.  
I would also like to thank D.~Dooling, R.~Easther, 
B.~Greene, M.~Grisaru and  M.~Jackson for helpful discussions 
and T.~H\"{u}bsch for enlightening correspondences.
%
%
This research was supported 
in part by NSERC (Canada) and FCAR (\Quebec).
   
\end{document}